\input{aipcheck}

\documentclass[
    ,final            
  ]
  {aipproc}

\layoutstyle{8x11double}

\begin{document}

\title{Searching for variable stars in Galactic Open Clusters}
\classification{<Replace this text with PACS numbers; choose from this list:
                \texttt{http://www.aip..org/pacs/index.html}>}

\author{Claudia Greco}{address={Geneva Observatory, University of Geneva, ch. des Maillettes 51, CH-1290, Sauverny, Switzerland}}
\author{Nami Mowlavi}{address={Geneva Observatory, University of Geneva, ch. des Maillettes 51, CH-1290, Sauverny, Switzerland}}
\author{Laurent Eyer}{address={Geneva Observatory, University of Geneva, ch. des Maillettes 51, CH-1290, Sauverny, Switzerland}}
\author{Maxime Spano}{address={Geneva Observatory, University of Geneva, ch. des Maillettes 51, CH-1290, Sauverny, Switzerland}}
\author{Mihaly Varadi}{address={Geneva Observatory, University of Geneva, ch. des Maillettes 51, CH-1290, Sauverny, Switzerland}}
\author{Gilbert Burki}{address={Geneva Observatory, University of Geneva, ch. des Maillettes 51, CH-1290, Sauverny, Switzerland}}

\begin{abstract}
%
%
A long-term project, aiming at systematic search for variable stars in Galactic Open Clusters, was started 
at the Geneva Observatory in 2002. We have been observing regularly a sample of twenty-seven Galactic 
Open Clusters in the  $U$, $B$, $V$ Geneva filters. The goal is to identify and to study their variable stars, as 
well as the connection between the variable stars in a cluster and the cluster properties. We present the status of 
this work in progress, and show preliminary results for one of these clusters, IC~4651.

\end{abstract}

\maketitle


\section{Introduction}
  All  stars belonging to an Open Cluster (OC) are assumed to share the same age, chemical abundances, 
  distance and reddening. The already listed 1200 OCs in the disk of our Galaxy (Lynga 1987; Dias et al. 2002)  
  span a large interval in age and galactocentric distances: they have been used 
  as an excellent tool to probe the structure and the evolution -both chemical and dynamical- of the 
  Milky Way disk.

Stellar variability, whether due to intrinsic properties or geometrical effects, affects the light curve 
of stars at many phases of their evolution, with specific variability properties for each phase. This, 
together with the variety of variable star content from one cluster to another, provides independent 
measurements for the physical parameters of OCs. 
Yet, the connection between the types of variable found and the properties 
of the OC is not well know so far. 
\section{Project}
  Long-term observations of a sample of twenty-seven Galactic OCs 
have been  carried out in the $U$, $B$ and $V$ Geneva filters (see Cherix et al, 2006). Twelve OCs
 in the Southern Hemisphere have been observed 
with Euler-Cam, mounted on the 1.2m Euler Swiss Telescope at La Silla, Chile and  fifteen OCs in the Northen Hemisphere 
OCs with Merope-Cam installed on the 1.2m Mercator Belgian telescope, in La Palma, Canary Islands, Spain. The field of view
of the two cameras (Euler: 11.5$^{\prime}$ $\times$ 11.5 $^{\prime}$; Mercator: 6.5$^{\prime}$ $\times$  6.5$^{\prime}$) is centered
on the cluster and contains cluster stars as well as Milky Way field stars.

The goal of the project is to map the whole variable star content in each cluster in our database.
Our sample includes metal-poor (e.g. ${\langle {\rm [Fe/H]\rangle}} =-0.52 \ dex$, NGC~2324) and metal-rich 
(e.g. ${\langle {\rm [Fe/H]\rangle}} = 0.10 \ dex$, IC~4651) clusters, and spans from very young 
(e.g. $log(age)=7.160$, NGC~3766) to older (e.g. $log(age)=9.325$, NGC~7789) ages. In Table 1, 
we summarize some properties for the OCs 
included in this project. Literature data are taken from  WEBDA, \url{http://www.univie.ac.at/webda/presentation.html}.
\begin{figure}
  \includegraphics[height=.35\textheight]{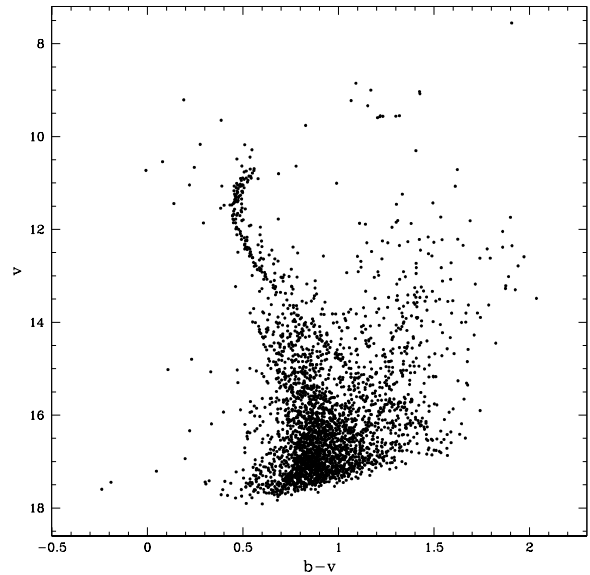}
  \caption{Color-Magnitude Diagram of IC~4651.}
\end{figure}

\begin{figure}
  \includegraphics[height=.52\textheight]{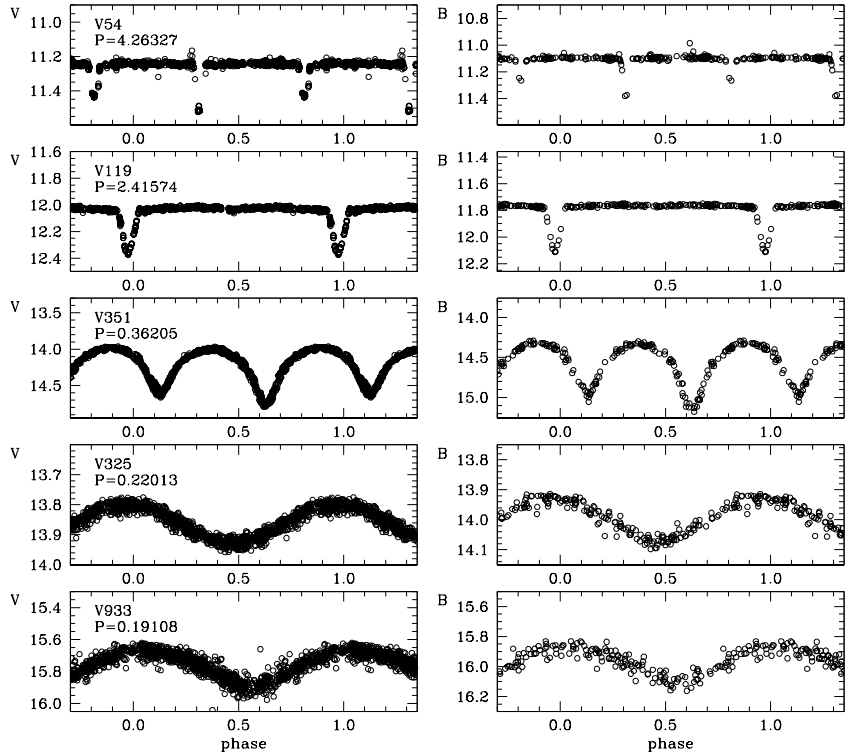}
   \caption{Folded light curves of some variable stars in the field of IC~4651. Periods are in days.}
\end{figure}

Data acquisition and data reduction are still in progress. For the Southern clusters we already collected more than 
200 $U$, 200 $B$ and 2000 $V$ images on a time baseline of 7 years.  A preliminary Color Magnitude Diagram (CMD),
 not yet calibrated to any 
standard system, is shown  in Fig.1 for the oldest cluster of our sample, IC~4651. 
Times series for Northern OCs include so far almost 1600 V image over a 5 year baseline.

\section{Data Reduction and Preliminary Results}

We  developed a semi-automatic pipeline to prereduce and reduce all the images. Such a long time baseline includes 
a lot of changes in the performance  of the telescope. To take these variations into account, the images have been 
reduced in groups according to the  period of collection. In the prereduction process, 
great care has been devoted to the analysis of the flatfield images and the pre/over scan values in order 
to infer the best correction for each image. 
We perform PSF photometry with the DaophotII/Allstar (Stetson 1998, 1999) packages. For each image, we build a 
proper PSF, using at least one hundred selected stars chosen among the brightest and most isolated ones. 
 All the reduced images are then aligned to a common reference list of stars in order to build the 
 CMD and the time 
series for each star. Time series are then searched for objects showing variability. 
 The time series of the candidate variables have been analyzed using two different codes. We used GrATis 
(Graphical Analyzer of Time Series; Di Fabrizio 1999; Clementini 2000) and Period04 (Lenz \& Breger 2005).
 The first variable stars found so far have periods spanning from 0.19  to 53 days, including 
\emph{bona fide} $\delta$ Scuti and binary stars. Several candidates show multi-periodic variability. 
In Fig. 2 we show folded light curves for some variable stars we identified.

\section{Conclusions}
 Given the large time baseline,  the project is well suited  for  long-term variability. An overall accuracy of a 
few mmag, plus  the long time baseline,  allows the  identification of various kinds of variable stars, 
including small period and small amplitude ones. 
 Preliminary analysis of the variable stars indeed confirms the potentialities of this project to identify 
 variable stars and to use them to characterize the properties of OCs.

\begin{table}
\begin{tabular}{lccccc}
\hline
\tablehead{1}{c}{b}{$Open Cluster$}
 & \tablehead{1}{c}{b}{$RA$}
 & \tablehead{1}{c}{b}{$DEC$} 
 & \tablehead{1}{c}{b}{$log(age)$}	
 & \tablehead{1}{c}{b}{${\langle {\rm [Fe/H]\rangle}}$\tablenote{dex}}	
 & \tablehead{1}{c}{b}{$Telescope$}	\\
\hline
\bf{NGC~3766}  & 11:36:14 & -61:36:30 &7.160 & -      & Euler \\
\bf{NGC~4103} & 12:06:40 & -61:15:00 &7.393 & -      & Euler \\
NGC~6755  & 19:07:49 & +04:16:00 &7.719 & - 	& Mercator \\
NGC~7654  & 23:24:48 & +61:35:36 &7.764 & -      & Mercator \\
NGC~7039  & 21:10:48 & +45:37:00 &7.280 & -      & Mercator \\
NCG~5617\tablenote{see Carrier 2009}  & 14:29:44 & -60:42:42 &7.915 & -      & Euler \\
NGC~6996  & 20:56:30 & -54:13:06 &8.000 & - 	& Euler \\
\bf{NGC6067}  & 16:13:11 & -54.13:06 &8.076 & 0.13   & Euler \\
\bf{NGC3247}  & 10:25:51 & -57:55:24 &8.083 & -      & Euler \\
NGC~2323  & 07:02:42 & -08:23:00 &8.096 & 0.02   & Euler \\
NGC~1513  & 04:09:57 & +49:30:54 &8.110 & -      & Mercator \\
NGC~6705  & 18:51:05 & -06:16:12 &8.302 & 0.13   & Mercator \\
\bf{NGC2437}  & 07:41:46 & -14:48:36 &8.390 & 0.05   & Euler \\
\bf{NGC3532}  & 11:05:39 & -58:45:12 &8.492 & -0.02  & Euler \\
NGC~2194  & 06:13:45 & +12:48:24 &8.515 & -	& Mercator \\
NGC~1907  & 05:28:05 & +35:19:30 &8.567 & -      & Mercator \\
\bf{NGC2447}  & 07:44:30 & -23:51:24 &8.588 & 0.03   & Euler \\
NGC~2324  & 07:04:07 & +01:02:42 &8.630 & -0.52  & Mercator \\
NGC~1245  & 03:14:41 & +47:14:12 &8.704 & 0.10   & Mercator \\
NGC~6811  & 19:37:17 & +46:23:18 &8.799 & - 	& Mercator \\
NGC~1901\tablenote{see Cherix 2006}  & 05:18:15 & -68:26:12 &8.920 & -	& Euler \\
NGC~6134	 & 16:27:46 & -49:09:06 &8.968 & 0.18   & Euler \\
NGC~2420  & 07:38:23 & +21:34:24 &9.048 & -0.26  & Mercator \\
\bf{IC~4651}   & 17:24:49 & -49:56:00 &9.057 & 0.10   & Euler \\
NGC~7789  & 23:57:24 & +56:42:30 &9.235 & -0.08  & Mercator \\
NGC~6939  & 20:31:30 & +60:39:42 &9.346 & 0.02   & Mercator \\
NGC~188	 & 00:47:28 & +85:15:18 &9.632 & -0.02  & Mercator \\
\hline
\end{tabular}
\caption{List of the twenty-seven OCs in the Geneva database. In bold, the eight clusters we have already reduced. References to
previous work from the same database are in footnotes. Literature data are taken from  WEBDA.}
\label{tab:a}
\end{table}

\begin{theacknowledgments}
We thank the Swiss National
Science Foundation for the continuous support to this project.
 \end{theacknowledgments}



\def\aj{AJ}%
\def\actaa{Acta Astron.}%
\def\araa{ARA\&A}%
\def\apj{ApJ}%
\def\apjl{ApJ}%
\def\apjs{ApJS}%
\def\ao{Appl.~Opt.}%
\def\apss{Ap\&SS}%
\def\aap{A\&A}%
\def\aapr{A\&A~Rev.}%
\def\aaps{A\&AS}%
\def\azh{AZh}%
\def\astrolet{Astronomy Letters}%
\def\baas{BAAS}%
\def\bac{Bull. astr. Inst. Czechosl.}%
\def\caa{Chinese Astron. Astrophys.}%
\def\cjaa{Chinese J. Astron. Astrophys.}%
\def\icarus{Icarus}%
\def\jcap{J. Cosmology Astropart. Phys.}%
\def\jrasc{JRASC}%
\def\mnras{MNRAS}%
\def\memras{MmRAS}%
\def\manual{ User's Manual for DAOPHOT II}
\def\na{New A}%
\def\nar{New A Rev.}%
\def\pasa{PASA}%
\def\pra{Phys.~Rev.~A}%
\def\prb{Phys.~Rev.~B}%
\def\prc{Phys.~Rev.~C}%
\def\prd{Phys.~Rev.~D}%
\def\pre{Phys.~Rev.~E}%
\def\prl{Phys.~Rev.~Lett.}%
\def\pasp{PASP}%
\def\pasj{PASJ}%
\def\qjras{QJRAS}%
\def\rmxaa{Rev. Mexicana Astron. Astrofis.}%
\def\skytel{S\&T}%
\def\solphys{Sol.~Phys.}%
\def\sovast{Soviet~Ast.}%
\def\ssr{Space~Sci.~Rev.}%
\def\zap{ZAp}%
\def\nat{Nature}%
\def\iaucirc{IAU~Circ.}%
\def\aplett{Astrophys.~Lett.}%
\def\apspr{Astrophys.~Space~Phys.~Res.}%
\def\bain{Bull.~Astron.~Inst.~Netherlands}%
\def\fcp{Fund.~Cosmic~Phys.}%
\def\gca{Geochim.~Cosmochim.~Acta}%
\def\grl{Geophys.~Res.~Lett.}%
\def\jcp{J.~Chem.~Phys.}%
\def\jgr{J.~Geophys.~Res.}%
\def\jqsrt{J.~Quant.~Spec.~Radiat.~Transf.}%
\def\memsai{Mem.~Soc.~Astron.~Italiana}%
\def\to{The Observatory}%
\def\mess{The Messenger}%
\def\newa{New Astronomy}%
\def\Nature{Nature}%
\def\nphysa{Nucl.~Phys.~A}%
\def\physrep{Phys.~Rep.}%
\def\physscr{Phys.~Scr}%
\def\planss{Planet.~Space~Sci.}%
\def\procspie{Proc.~SPIE}%
\def\Science{Science}%

\bibliographystyle{aipproc}   


\IfFileExists{\jobname.bbl}{}
 {\typeout{}
  \typeout{******************************************}
  \typeout{** Please run "bibtex \jobname" to optain}
  \typeout{** the bibliography and then re-run LaTeX}
  \typeout{** twice to fix the references!}
  \typeout{******************************************}
  \typeout{}
 }

\hyphenation{Post-Script Sprin-ger}

\end{document}